\documentclass[12pt,a4paper]{article}
\usepackage{latexsym}
\usepackage{amssymb}
\usepackage{amsmath}
\usepackage{amsfonts}
\usepackage{mathrsfs}
\usepackage{cite}

\catcode`@=11
\renewcommand{\section}{\@startsection{section}{1}{0pt}{\medskipamount}
{\medskipamount}{\large\bf}} \numberwithin{equation}{section}
\catcode`@=12

\newcommand{\be}{\begin{equation}}
\newcommand{\ee}{\end{equation}}

\newcommand{\lb}{\label}

\def\t{\theta}
\def\tr{{\rm tr}}
\def\tr{{\rm tr}\,}
\def\Tr{{\rm Tr}\,}
\def\cN{{\cal N}}

\def\cA{{\cal A}}

\def\bea{\begin{eqnarray}}
\def\eea{\end{eqnarray}}
\def\nn{\nonumber}

\def\cN{{\cal N}}

\def\f{\frac}

\def\tr{{\rm tr}\,}

\def\nn{\nonumber}

\def\d{\delta}

\def\g{\gamma}

\def\ve{\varepsilon}

\def\sB{\stackrel{\frown}{\square}}

\def\eq{\eqref}
\def\pr{\partial}
\def\nb{\nabla}

\numberwithin{equation}{section}

\date{\it  }
\begin{document}
\begin{titlepage}

\begin{center}
\vspace{1cm} {\Large\bf One-loop divergences in $6D$, $\cN=
(1,0)$ SYM theory}
\vspace{1.5cm}

 {\bf
 I.L. Buchbinder\footnote{joseph@tspu.edu.ru }$^{\,a,b}$,
 E.A. Ivanov\footnote{eivanov@theor.jinr.ru}$^{\,c}$,
 B.S. Merzlikin\footnote{merzlikin@tspu.edu.ru}$^{\,a,d}$,
 K.V. Stepanyantz\footnote{stepan@m9com.ru}$^{\,e}$
 }
\vspace{0.4cm}

 {\it $^a$ Department of Theoretical Physics, Tomsk State Pedagogical
 University,\\ 634061, Tomsk,  Russia \\ \vskip 0.15cm
 $^b$ National Research Tomsk State University, 634050, Tomsk, Russia \\ \vskip 0.1cm
 $^c$ Bogoliubov Laboratory of Theoretical Physics, JINR, 141980 Dubna, Moscow region,
 Russia \\ \vskip 0.1cm
 $^d$ Department of Higher Mathematics and Mathematical Physics,\\
\it Tomsk Polytechnic University, 634050, Tomsk, Russia\\
\vskip 0.1cm
 $^e$ Department of Theoretical Physics, Moscow State University,
119991, Moscow, Russia

}
\end{center}
\vspace{0.4cm}

\begin{abstract}
We consider, in the harmonic superspace approach, the six-dimensional $\cN=(1,0)$  supersymmetric
Yang-Mills gauge multiplet minimally coupled to a hypermultiplet in an arbitrary representation of the gauge group.
Using the superfield proper-time and background-field techniques, we compute the divergent
part of the one-loop effective action depending on both the gauge multiplet and the hypermultiplet.
We demonstrate that in the particular case of $\cN=(1,1)$ SYM theory, which corresponds
to the hypermultiplet in the adjoint representation, all one-loop divergencies vanish,
so that $\cN=(1,1)$ SYM theory is one-loop finite {\it off shell}.

\end{abstract}
\end{titlepage}

\setcounter{footnote}{0}

\setcounter{page}{1}

\section{Introduction}
\hspace*{\parindent}

The ultraviolet behavior of extended supersymmetric Yang-Mills (SYM)
theories in higher dimensions ($D \geq 5$) represents an exciting
subject with the long history \cite{HS,HS1,BHS,BHS1,FT}. In this
work we focus on the $6D$ SYM theory coupled to hypermultiplets. We
formulate the theory in $6D$, $\cN=(1,0)$ harmonic superspace
\cite{GIKOS,GIOS,HST,HSW,Z,ISZ} and develop the corresponding
background superfield method. As the basic topic, we expose the
ultraviolet properties of the one-loop effective action for this
theory in the general case when the hypermultiplet lies in an
arbitrary representation of the gauge group. In the particular case
of the adjoint representation, the considered ${\cal N}=(1,0)$ SYM -
hypermultiplet system amounts to $6D$, $\cN=(1,1)$ SYM theory
formulated in terms of $\cN=(1,0)$ harmonic superfields.

In a recent work \cite{BIMS} we have calculated the divergent part of the one-loop effective
action for the abelian $6D$, $\cN=(1,0)$ gauge theory, in which the vector (gauge) multiplet
interacts with a hypermultiplet. The basic tools were the background superfield method
and proper time technique appropriately adapted to $6D$, $\cN=(1,0)$  harmonic superspace.
By explicit calculations we confirmed the general structure of one-loop counterterms
which was analyzed earlier in refs. \cite{HS,BIS} on the pure symmetry grounds.
In the present paper we generalize this study to the non-abelian case.
We consider the $6D$, $\cN=(1,0)$ model in which the SYM multiplet interacts
with the hypermultiplet in an arbitrary representation of gauge group,
the adjoint and fundamental representations being particular cases. We extend the background superfield
method to this general case of $6D$, $\cN=(1,0)$ SYM theory with
the hypermultiplet matter. In many aspects, it is similar to the well-developed
background superfield method for $4D, {\cal N}=2$ SYM theory
with hypermultiplets \cite{BUCH1,BUCH2}. Using the $6D$, $\cN=(1,0)$
harmonic background superfield method constructed and the proper time technique,
we calculate the divergent part of the one-loop effective action
in the considered $6D$, $\cN=(1,0)$ model. It should be emphasized that
we take into account the full set of contributions depending
on both the background gauge multiplet and the hypermultiplet.
To the best of our knowledge, the  explicit calculation
of the hypermultiplet-dependent divergent contributions
to effective action of $6D$ SYM theories has never been accomplished earlier,
and it is the pivotal point of our consideration.

It is well known that both $6D$, $\cN=(1,0)$ and $6D$, $\cN=(1,1)$ SYM theories
at one loop are on-shell finite \cite{HS,BIS}. For the $6D$, $\cN=(1,0)$ theory without hypermultiplets
this result is easily recovered from the quantum calculations. The main result of the present
work is the explicit proof of the absence of one-loop logarithmic divergencies in $6D$, $\cN=(1,1)$ SYM theory {\it off shell}.
We demonstrate this by calculating the divergent part of the one-loop effective action in $\cN=(1,1)$ SYM
theory formulated in terms of $\cN=(1,0)$ harmonic gauge and hypermultiplet superfields,
both in the adjoint representation of the gauge group  \cite{BIS}.
We start with a general $6D$, $\cN=(1,0)$ SYM - hypermultiplet action and find
the one-loop contributions to the divergent part of the effective action.
We demonstrate that the numerical factors depending on the gauge group and on the representation
of the hypermultiplet vanish in the case when the hypermultiplet is in the adjoint representation
of the gauge group. Hence, for the $\cN=(1,1)$ SYM theory we establish the absence of logarithmic divergencies
in the one-loop effective action. The similar phenomenon takes place in $\cN=4$ SYM theory
in four dimensions formulated in terms of $\cN=2$ superfields (see, e.g., \cite{BUCH2}).

It should be pointed out that in a certain sense the {\it off-shell} absence
of the one-loop divergencies in that part of the total $\cN=(1,1)$ SYM effective action
which depends only on gauge background superfields is an expected result.
It is dictated by the formal structure of this one-loop effective action, in which the contributions from
the ghost superfields are canceled by the corresponding contribution from quantum hypermultiplet
in the adjoint representation. Once again, this happens in the full analogy with $4D$, $\cN=4$ case \cite{BUCH1}.
However, taking the background hypermultiplet parts of the one-loop effective action into account entails
a few technical problems. The basic one is that, after making the background-quantum splitting,
we encounter the mixed terms involving the quantum gauge superfields along with the hypermultiplet ones.
In order to diagonalize the action, we are led to make a non-local shift of hypermultiplet
variables \cite{Kuz07,BM15,OV} which induces an additional background hypermultiplet
dependence in the one-loop effective action caused by the contributions from the quantum gauge multiplet.

The paper is organized as follows. In section 2 we briefly outline the gauge theory
in  $6D, \,{\cal N}=(1,0)$ harmonic superspace and fix our $6D$ notations and conventions.
Section 3 presents the harmonic superspace background superfield method for $\cN=(1,0)$ SYM theory.
In section 4 we perform the direct calculations of the one-loop divergences in the model under consideration.
In section 5 we  summarize the results and discuss the problems for further study.


\section{Gauge theory in $6D$, $\cN=(1,0)$ harmonic superspace}
Our consideration in this section (including notations, conventions
and terminology) will closely follow ref. \cite{BIS}.

The $6D, {\cal N}=(1,0)$ gauge covariant derivatives in the ``central basis'' are defined by
 \be
 \nb_{\cal M}=D_{\cal M}+i\cA_{\cal M},
 \ee
where $D_{\cal M} = (D_M, D^i_a)$ are the flat derivatives. Here
$M=0,..,5,$ is the $6D$ vector index and $a=1,..4,$ is the spinorial
one. The superfield $\cA_{\cal M}$ is the gauge superconnection.
The covariant derivatives transform under the gauge group as
 \be
 \nb'_{\cal M}=e^{{\rm i}\tau}\nb_{\cal M} e^{-{\rm i}\tau}\,, \quad
 \tau^{+}=\tau\,.
 \label{tau}
 \ee

The fundamental object of $6D, {\cal N}=(1,0)$ SYM theory is revealed after extending
the standard  $6D, {\cal N}=(1,0)$ superspace $z :=(x^M, \theta^a_i)$
by $SU(2)$ harmonics $u^\pm_i, \, u^{+i}u^-_i =1\,,$ and singling out,
in this extended harmonic $6D, {\cal N}=(1,0)$ superspace $(z, u)$,
an analytic subspace $(\zeta, u)$ containing four independent Grassmann coordinates along with the harmonics $u^\pm_i$.
All geometric quantities of the theory
are expressed in terms of the hermitian analytic gauge connection $V^{++}(\zeta, u) = \widetilde{V^{++}(\zeta, u)}\,,$
 \bea
V^{++}=(V^{++})^A T^A\,, \qquad (T^A)^+ = T^A\,, \label{Vfirst}
 \eea
where the generalized conjugation $\widetilde{\;}$ is defined in \cite{GIOS} and $T^A$ are the generators of the gauge group.

For simplicity, we will consider only simple gauge groups. In our notation
the generators of the fundamental representation $T^A_{\mbox{\scriptsize f}}\equiv t^A$
are normalized by the condition $\mbox{tr}(t^A t^B) = \frac{1}{2}\delta^{AB}$. For an arbitrary representation $R$,
which can be in general reducible,
 \bea\label{Casimirs}
[T^A, T^B] = i f^{ABC} T^C\,, \qquad
\tr(T^A T^B) = T(R) \d^{AB}\,, \qquad (T^A)_m{}^{l} (T^A)_l{}^{n} = C(R)_m{}^{n}.
 \eea

\noindent
If  $R$ is irreducible, we obtain:

\begin{equation}\label{For_Irrep}
C(R)_m{}^n = C_2(R) \delta_m^n, \quad C_2(R) = T(R) \frac{d_G}{d_R}\,,
\end{equation}

\noindent
where $C_2(R)$ is the second Casimir for the representation $R$,  $d_G\equiv \delta_{AA}$
is the dimension of the gauge group, and $d_R \equiv \delta_m^m$ is the dimension of
the irreducible representation $R$. In the case when $R$ is a reducible representation,
$R = \sum_i R_{(i)}$, we have (in the matrix notation)
\begin{equation}
T(R) = \sum_i T(R_{(i)}) d_{R_{(i)}}\,, \quad C(R)= \sum_i C_2 (R_{(i)}) I_{(i)}\,, \quad d_{R_{(i)}} = \tr I_{(i)}\,,
\end{equation}
whence
$$
T(R_{(i)}) = C_2 (R_{(i)})\,\frac{d_{R_{(i)}}}{d_G}\,.
$$
For the adjoint representation the generators are written as $(T^C_{\mbox{\scriptsize Adj}})_A{}^B =i f^{ACB}$. Consequently,

\begin{equation}
T(\mbox{Adj})=C_2, \qquad  C(\mbox{Adj})_m{}^{n}=C_2\delta_m^{n}.
\end{equation}

 The connection $V^{++}\,$, \eq{Vfirst}, covariantizes the flat analyticity-preserving harmonic derivative $D^{++}$:
\bea
D^{++} \;\Rightarrow \; \nabla^{++} = D^{++} + i V^{++}\,, \quad (V^{++})' = -i e^{i\lambda^A T^A}D^{++} e^{-i\lambda^A T^A}
 +e^{i\lambda^A T^A} V^{++} e^{-i\lambda^A T^A}\,,\lb{V++transf}
\eea
where $\lambda^A(\zeta, u) = \widetilde{\lambda^A(\zeta, u)}$ is the real gauge group parameter in the ``$\lambda$-basis''.
Another important object is the non-analytic
harmonic connection $V^{--}= (V^{--})^A T^A$ covariantizing the flat derivative $D^{--}$
\bea
D^{--} \;\Rightarrow \; \nabla^{--} = D^{--} + i V^{--}\,, \quad (V^{--})' = -i e^{i\lambda^A T^A}D^{--} e^{-i\lambda^A T^A}
 +e^{i\lambda^A T^A} V^{--} e^{-i\lambda^A T^A}\,.
\eea
It is not independent and is related to $V^{++}$ by the harmonic flatness condition
\bea
[\nabla^{++}, \nabla^{--}] = D^0\;\Leftrightarrow \;D^{++} V^{--} - D^{--}V^{++} + i [V^{++},V^{--}]=0\,, \label{zeroc}
\eea
where $D^0$ is the operator counting the harmonic $U(1)$ charges of the involved superfields. The formal solution of (\ref{zeroc}) is
 \be
V^{--}(z,u)=  \sum\limits^{\infty}_{n=1} (-i)^{n+1  } \int du_1\ldots
du_n\, \frac{V^{++}(z,u_1 )\ldots V^{++}(z,u_n )}{(u^+ u^+_1)(u^+_1
u^+_2)\ldots (u^+_n u^+)}\,.
 \ee
Using the zero curvature condition \eq{zeroc}, one can derive a
useful relation between arbitrary variations of harmonic connections
\cite{BIS}
 \be
\d V^{--} = \f12(\nb^{--})^2\d V^{++} - \f12 \nb^{++}(\nb^{--}\delta
V^{--}) \,. \label{var}
 \ee

All the geometric quantities of the theory are expressed in terms of $V^{--}$.
The covariant derivatives in the $\lambda$-frame can be written as
 \bea
 \nb^+_a = D^+_a\,, \quad \nb^-_a = D^-_a + i {\cal A}^-_a\,,
 \quad \nb_{ab} = \pr_{ab} + i {\cal A}_{ab}\,,
 \eea
where superfield connections are determined as
 \bea
 {\cal A}^-_a= i D^+_a V^{--}\,, \qquad {\cal A}_{ab} = \f12
 D^+_a D^+_b V^{--} \,.
 \eea
The covariant derivatives satisfy the algebra
 \begin{equation}
\{\nb_a^+,\nb^-_b\}=2i\nb_{ab},\qquad
[\nb_c^\pm,\nb_{ab}]=\f{i}2\ve_{abcd}W^{\pm\, d},\qquad [\nb_M, \nb_N] = i F_{MN}\,, \label{alg2}
 \end{equation}
where $\nb_{ab} = \f12(\g^M)_{ab} \nb_M$ and $W^{a\,\pm}$ is the covariant superfield strength
 \bea
W^{+a}=-\frac{1}{6}\varepsilon^{abcd}D^+_b D^+_c D^+_d
V^{--}\,, \quad W^{-a} = \nb^{--}W^{+a}\,.\label{W+Def}
 \eea
We also  define the Grassmann-analytic superfield \cite{BIS}

\begin{equation}\label{identity00}
F^{++} \equiv \frac14 D^+_a W^{+ a} =(D^+)^4 V^{--}\,,
\end{equation}

\noindent
such that

\begin{equation}
 \quad D^+_a W^{+ b} = \delta^b_a F^{++}\,, \quad D^+_aF^{++} = 0\,, \;\nb^{++} F^{++}=0\,.
\label{identity0}
\end{equation}

\noindent
It will be used for constructing the background field formalism and counterterms
in the next sections.

The harmonic covariant derivatives $\nb^{\pm\pm} = D^{\pm\pm} + i
V^{\pm\pm}$ act on the arbitrary analytic superfields ${\cal F}$ in an arbitrary representation of the gauge group
as
 \bea
 (\nb^{\pm\pm} {\cal F})_m =
 \Big( D^{\pm\pm} \d_m{}^{n} +i (V^{\pm\pm})^C (T^C)_m{}^{n} \Big){\cal F}_{n}\,
 \equiv (\nb^{\pm\pm})_m{}^{n} {\cal F}_{n}\,. \label{nbRq}
 \eea
If ${\cal F}$ belongs to the adjoint representation, then the above equation gives
 \bea
 (\nb^{\pm\pm} {\cal F})^A = \Big( D^{\pm\pm} \d^{AB} - f^{ACB} (V^{\pm\pm})^C
 \Big){\cal F}^B\, \equiv (\nb^{\pm\pm})^{AB} {\cal F}^B\,. \label{nbAdj}
 \eea

The superfield action of $6D,\,\cN=(1,0)$ SYM interacting with
a hypermultiplet  has the form
 \bea
 \label{S0}S_0[V^{++}, q^+]&=&
\frac{1}{f^2}\sum\limits^{\infty}_{n=2} \frac{(-i)^{n}}{n} \tr \int
d^{14}z\, du_1\ldots du_n \frac{V^{++}(z,u_1 ) \ldots
V^{++}(z,u_n ) }{(u^+_1 u^+_2)\ldots (u^+_n u^+_1 )}  \nn \\
&& - \int d\zeta^{(-4)} du\, \tilde{q}^{+\,m} (\nabla^{++})_m{}^{n}
q^{+}_{n}\,, \label{S0}
 \eea
where $f$ is a dimensionful coupling constant ($[f]=-1$). In the SYM part of this action $V^{++} = V^{++A} t^A$
with $t^A$ being generators of the fundamental representation, while in the hypermultiplet
part of the action $(V^{++})_m{}^n = V^{++ A} (T^A)_m{}^n$, where $T^A$ are generators of the representation
for the hypermultiplet. The
action \eq{S0} is invariant under the gauge transformation \eq{V++transf} and
 \be
( q^{+}_{m})' =  (e^{i\lambda^A T^A })_m{}^{n}  q^{+}_{n}\,.
 \label{gtr}
 \ee

Classical equations of motion following from the action \eq{S0} read
 \bea
 \frac{{\delta}S}{{\delta}(V^{++})^A}=0 \,&\Rightarrow& \,
 \frac{1}{f^2} (F^{++})^A + i\tilde{q}^{+\,m} \, (T^A)_m{}^{n}\, q^{+}_{n} = 0\,, \lb{eqmF}
 \\
 \frac{{\delta}S}{{\delta}\tilde{q}^{+\, m}} = 0 \,&\Rightarrow& \,(\nabla^{++})_m{}^{n}q^{+}_{n}
 = 0\,. \label{eqm}
  \eea
The $\;\widetilde{\;}\;$ - reality of Eq. \eq{eqmF} (as well as  of the action \eq{S0}) is guaranteed by the conjugation
rules $\widetilde{{\tilde q}^{+}} = - q^+\,, \; \widetilde{F^{++}} = F^{++}\,$ \cite{GIOS}.

\section{Background field formalism for $\cN=(1,0)$ SYM theory}
In the present paper we generalize the background field method
developed in \cite{BIMS} for the abelian case to the non-abelian model \eqref{S0}. The construction of gauge
invariant effective action in the model under consideration is very
similar to that in $4D, \cN=2$ supersymmetric gauge theories
\cite{BUCH1}, \cite{BP07} (see also the reviews \cite{BUCH2}).\footnote{The background field method can be also constructed
in the ordinary ${\cal N}=2$ superspace \cite{Howe}. However, this approach encounters a problem
of an infinite number of the Faddeev--Popov ghosts.}

One splits the superfields $V^{++}, q^{+}$ into the sum of the
``background'' superfields $V^{++}, Q^{+}$ and the ``quantum'' ones
$v^{++}, q^{+}\,$,
 \be
 V^{++}\to V^{++} + fv^{++}, \qquad q^{+} \to Q^{+} + q^{+}\,,
 \ee
and then expand the action in a power series in quantum fields. As a
result, we obtain the classical action as a functional of background
superfields and quantum superfields.  The original infinitesimal
gauge transformations are realized in two different ways:  as the
\emph{background} transformations:
\begin{equation}\label{cltr}
\delta V^{++}=-\nb^{++}\lambda, \quad \delta
v^{++}=-i[v^{++},\lambda]~,
\end{equation}
and as the \emph{quantum} transformations\footnote{We denote the parameters of these transformations
by the same letter, hoping that this will not lead to confusion.}
\begin{equation}\label{qutr}
\delta V^{++}=0,\quad  \delta v^{++}=-\nb^{++}\lambda -i[v^{++},
\lambda].
\end{equation}

To construct the gauge invariant effective action, we need to impose
the gauge-fixing conditions only on quantum superfields. We
introduce the gauge-fixing function in the full analogy
with $4D$ case \cite{BUCH1,BUCH2}
 \be
\label{gf} {\cal F}^{(+4)}_\tau=D^{++}v^{++}_\tau
=e^{-ib}(\nb^{++}v^{++})e^{ib}=e^{-ib}{\cal F}^{(+4)}e^{ib}~,
 \ee
where $b(z)$ is a background-dependent gauge bridge superfield  and
$\tau$ means $\tau$-frame (see, e.g., \cite{GIOS}). We consider the
non-abelian gauge theory, where the gauge-fixing function \eqref{gf}
is background-dependent. The gauge-fixing function transforms
according to the law
\begin{equation}\label{trf}
\delta{\cal F}^{(+4)}_\tau=-e^{-ib}\{\nb^{++}(\nb^{++}\lambda
+i[v^{++},\lambda])\}e^{ib}
\end{equation}
under the quantum transformations (\ref{qutr}). Eq. (\ref{trf})
leads to the Faddeev-Popov determinant
$$\Delta_{FP}[v^{++},V^{++}]=\mbox{Det}(\nb^{++}(\nb^{++}
+iv^{++}))\,.$$ Following the standard procedure, we can obtain a
path-integral representation for $\Delta_{FP}[v^{++},V^{++}]$ by
introducing two real analytic fermionic ghosts ${\bf b}$ and ${\bf
c}$, both in the adjoint representation of the gauge group. The
corresponding ghost action is
\begin{equation}\label{FP}
S_{FP}[{\bf b}, {\bf c}, v^{++}, V^{++}] =\tr\int
d\zeta^{(-4)}du\,\, {\bf b}\nb^{++}(\nb^{++}{\bf c} +i[v^{++},{\bf
c}]).
\end{equation}

As a result, we arrive at the effective action $\Gamma[V^{++}, Q^+]$
in the form
 \bea \label{path}
 e^{i\Gamma[V^{++},Q^+]}=\int {\cal D}v^{++}{\cal D}q^+ {\cal D}{\bf b}{\cal D}{\bf
 c}\,
\delta[{\cal F}^{(+4)}-f^{(+4)}]\, e^{i \big\{ S_0[V^{++} + fv^{++},
Q^+ + q^+] + S_{FP}[{\bf b}, {\bf c}, v^{++}, V^{++}]\big\}}\,,
 \eea
where $f^{(+4)}(\zeta,u)$ is an external Lie-algebra valued analytic
superfield which is independent of $V^{++}$, and $\delta[{\cal
F}^{(+4)}-f^{(+4)}]$ is the  functional analytic delta-function. As
the next step,  we average the right-hand side in Eq. (\ref{path}) with
the weight
\begin{equation}\label{weight}
\Delta[V^{++}]\exp\Big\{\frac{i}{2}\mbox{tr}\int d^{14}z du_1du_2
f_\tau^{(+4)}(z,u_1)\frac{(u^-_1u^-_2)}{(u^+_1u^+_2)^3}f_\tau^{(+4)}(z,u_2)\Big\}\,.
\end{equation}
Following the Faddeev-Popov method, the functional $\Delta[V^{++}]$
is determined from the equation
\begin{equation}
1=\Delta[V^{++}]\int {\cal
D}f^{(+4)}\exp\Big\{\frac{i}{2}\mbox{tr}\int d^{14}z du_1du_2\,
f_\tau^{(+4)}(z,u_1)\frac{(u^-_1u^-_2)}{(u^+_1u^+_2)^3}f_\tau^{(+4)}(z,u_2)\Big\}\,.
\label{3.9}
\end{equation}
Passing in this expression to the analytic subspace, we obtain
 \bea
\Delta^{-1}[V^{++}]&=&\int {\cal
D}f^{(+4)}\exp\Big\{\frac{i}{2}\mbox{tr}\int d\zeta^{(-4)}_1
d\zeta^{(-4)}_2 du_1du_2\,
f^{(+4)}(\zeta_1,u_1)A(1,2)f^{(+4)}(\zeta_2,u_2)\Big\} \nn \\
&=&\mbox{Det}^{-1/2}A\,. \label{3.10}
 \eea
Here, like in $4D$ case \cite{BUCH1,BUCH2}, we have introduced the
special background-dependent operator $A$, which arose when we
passed from \eq{3.9} to \eq{3.10}. This operator depends on the
background field through a background-dependent bridge $b(z)$ and
has the form
\begin{equation}
A(1,2) = \frac{(u_1^- u_2^-)}{(u_1^+ u_2^+)^3} (D_1^+)^4 (D_2^+)^4
\Big[(e^{ib_1} e^{-ib_2})_{\rm Adj} \delta^{14}(z_1-z_2)\Big]\,,
\end{equation}
where
 \bea
 (e^{ib_1}e^{-ib_2})_{\rm Adj} f^{(+4)}(\zeta_2,u_2) = e^{ib_1}e^{-ib_2} f^{(+4)}(\zeta_2,u_2) e^{ib_2}
e^{-ib_1}\,.
 \eea

We note that operator $A(1,2)$ acts in the space of analytic
superfields, which take values in the Lie algebra of the gauge
group. Thus, we have derived the following formal expression for the
functional $\Delta[V^{++}]$
\begin{equation}\label{Det}
\Delta[V^{++}]=\mbox{Det}^{1/2}A\,.
\end{equation}

To calculate the functional determinant for the operator $A$, we do
not need the explicit form for it.  We represent the determinant for
this operator through a functional integral over analytic
superfields,
\begin{equation}\label{Det1}
\mbox{Det}^{-1}A=\int {\cal D}\chi^{(+4)}{\cal
D}\rho^{(+4)}\exp\Big\{i\tr \int
d\zeta^{(-4)}_1du_1d\zeta^{(-4)}_2du_2\,\, \chi^{(+4)}(1)
A(1,2)\rho^{(+4)}(2)\Big\},
\end{equation}
and, as in $4D$ case, make use of the following substitution of the functional
variables
 \bea
 \rho^{(+4)}=(\nb^{++})^2\sigma, \quad
\mbox{Det}\left(\frac{\delta\rho^{(+4)}}{\delta\sigma}\right)=\mbox{Det}(\nb^{++})^2~.
 \eea
Then we find (see a similar calculation in \cite{BUCH1,BUCH2})
 \bea
 \label{Det2} \mbox{Det}^{-1}A = \mbox{Det}(\nb^{++})^2
 \int {\cal D}\chi^{(+4)}{\cal D}\sigma\,\exp\Big\{i\tr\int d\zeta^{(-4)}du\,
\chi^{(+4)}\sB_\lambda \sigma\Big\}\,.
 \eea
Here,  the operator $\sB_\lambda$ is the covariant d'Alembertian.
Hereafter we use the formal definition for this covariant
d'Alembertian $\sB_\lambda$ in $\lambda$-frame
\begin{equation}\label{smile}
\sB_\lambda=\frac{1}{2}(D^+)^4(\nb^{--})^2\,.
\end{equation}
It is possible to present this operator as a sum of two terms,
\begin{equation}\label{Box_As_A_Sum}
\sB_\lambda = \sB + X,
\end{equation}
where
\begin{eqnarray}\label{Box_First_Part}
&& \sB = \eta^{MN} \nabla_M \nabla_N + W^{+a} \nabla^{-}_a + F^{++} \nabla^{--} - \frac{1}{2}(\nabla^{--} F^{++})\,,\\
\label{Box_Second_Part} && X = \Big(W^{-a} - W^{+a} \nabla^{--} + 2i
\nabla^{ab} \nabla^{-}_b\Big) D_a^+ + \Big(i\nabla^{ab} \nabla^{--}
- \frac{1}{4}\varepsilon^{abcd} \nabla^{-}_c \nabla^{-}_d\Big) D_a^+ D_b^+\qquad\nonumber\\
&& - \nabla^{--} \nabla^{-}_d (D^+)^{3d} + \frac{1}{2} (\nabla^{--})^2 (D^+)^4.\label{Box_Rest}
\end{eqnarray}
In this equation we use the notation
\begin{equation}
(D^+)^{3d}\equiv -\frac{1}{6}\varepsilon^{dabc} D^+_a D^+_b D^+_c;
\qquad \nabla^{ab}\equiv \frac{1}{2} \varepsilon^{abcd} \nabla_{cd}.
\end{equation}
The presentation (\ref{Box_As_A_Sum}) is convenient, because the
operator $X$ gives vanishing contribution acting on the analytic
superfields. Therefore, when acting on the analytic superfields, the
operator $\sB_\lambda$ is reduced to the operator $\sB$.

In every case we should determine the space of superfields on which
the operator \eq{smile} acts,  namely, the harmonic $U(1)$ charge of
superfield and the representation of gauge group to which it
belongs. Using Eqs. (\ref{Det})-(\ref{Det2}), one obtains
\begin{equation}
\Delta[V^{++}]=\mbox{Det}^{-1/2}(\nb^{++})^2
\mbox{Det}^{1/2}\sB~.
\end{equation}
Finally, we can represent the functional determinant $\Delta[V^{++}]$
as the functional integral over bosonic real analytic
superfield $\varphi$ taking values in the Lie algebra of the gauge
group,
 \bea\label{NK}
\Delta[V^{++}]&=&\mbox{Det}^{1/2}{\sB}\int {\cal D}\varphi
\exp\Big\{ i S_{NK}[\varphi,V^{++}]\Big\}\,, \\
S_{NK} &=& \frac{1}{2}\tr \int d\zeta^{(-4)}du\, \varphi
(\nb^{++})^2\varphi\,.
 \eea
Like in $4D$ case, $\varphi$ is the Nielsen-Kallosh ghost. As a result, we see that the $6D$, $\cN=(1,0)$ SYM theory,
in the close analogy  with
$4D$, ${\cal N}=2$ SYM, in the background field approach is described by the three ghosts: two fermionic ghosts ${\bf b}$ and
${\bf c}$ together with the single bosonic ghost $\varphi$.

According to \eq{gf}, the gauge-fixing part of the quantum field
action has the form
  \be
S_{GF}[v^{++}, V^{++}] = -\frac{1}{2}\tr \int d^{14}z du_1
du_2\,\frac{v_\tau^{++}(1)v_\tau^{++}(2)}{(u^+_1u^+_2)^2}
 + \frac{1}{4}\tr \int
d^{14}z du\, v_\tau^{++} (D^{--})^2 v_\tau^{++}\,. \label{SGF}
 \ee
The action \eq{SGF} depends on the background field $V^{++}$ through
the background gauge bridge $b$, $v_\tau^{++} = e^{-ib} v^{++} e^{ib}$.

Summarizing, one can write the final expression for the effective action
\eq{path} as follows
 \bea
 e^{i \Gamma[V^{++},Q^+]} = \mbox{Det}^{1/2}\sB \int {\cal
D}v^{++}\,{\cal D}q^+\, {\cal D}{\bf b}\,{\cal D}{\bf c}\,{\cal
D}\varphi\,\, e^{iS_{quant}[v^{++}, q^+, {\bf b}, {\bf c}, \varphi,
V^{++}, Q^+]}.\label{path2}
 \eea
Here, the quantum action $S_{quant}$ has the structure
 \bea
S_{quant} &=& S_{0}[V^{++}+fv^{++}, Q^+ + q^+] + S_{GF}[v^{++},
V^{++}] \nn \\ &&+ S_{FP}[{\bf b}, {\bf c}, v^{++}, V^{++}] +
S_{NK}[\varphi, V^{++}].
 \eea

In the one-loop approximation, the first quantum correction to the
classical action, $\Gamma^{(1)}[V^{++}, \; Q^+]\,$, is given by the
following path integral \cite{BUCH1,BP07}:
 \be
 e^{ i\Gamma^{(1)}[V^{++}, Q^+]} =\mbox{Det}^{1/2}\sB \int {\cal
D}v^{++}\,{\cal D}q^+\, {\cal D}{\bf b}\,{\cal D}{\bf c}\,{\cal
D}\varphi\,\,\, e^{iS_{2}[v^{++}, q^+, {\bf b}, {\bf c}, \varphi,
V^{++}, Q^+]}\,.
 \label{Gamma0}
 \ee
In this expression, the full quadratic action $S_2$ is the sum of three terms. These are the classical
action \eq{S0} in which the  background-quantum splitting was performed,
the gauge-fixing action \eqref{SGF} and the actions for the ghost
superfields \eq{FP} and \eq{NK}:
  \bea
 S_2 &=& \frac{1}{2}\int d\zeta^{(-4)}du\, v^{++\,A}\sB^{AB} v^{++\,B}
 + \int d\zeta^{(-4)}du\,{\bf b}^A(\nb^{++})^{2\,AB}{\bf c}^B \nn \\
 &&  +\frac{1}{2}\int d\zeta^{(-4)}du\,\varphi^A(\nb^{++})^{2\,AB}\varphi^B
 - \int d\zeta^{(-4)}du\, \tilde{q}^{+\,m}
 (\nb^{++})_m{}^{n}q^{+}_{n} \nn \\
 &&- \int d\zeta^{(-4)}du\Big\{
  \widetilde{Q}^{+\,m} if(v^{++})^C (T^C)_m{}^{n} q^{+}_{n}+\tilde{q}^{+\,m} if(v^{++})^C (T^C)_m{}^{n}
Q^{+}_{n}\Big\}. \label{S2}
 \eea
Hereafter, we write all the group indices explicitly. The operator
$\sB$ \eq{smile} transforms the analytic superfields $v^{++}$ into
analytic superfields and, according to \eq{nbAdj}, has the following
structure
 \bea
  \sB^{AB} &=& \f12 (D^+)^4\Big\{(D^{--})^2 \d^{AB} - 2 f^{ACB}(V^{--})^C D^{--}
 -f^{ACB}(D^{--} V^{--})^C \nn \\
 &&\qquad\qquad+ f^{ACE} f^{EDB}(V^{--})^C(V^{--})^D\Big\}. \label{box1}
 \eea
The Green function, associated with \eq{box1}, {\it i.e.}
$G^{AB}_{(2,2)}(z_1,u_1|z_2,u_2) = i \langle0|{\rm T} (v_1^{++})^A
(v_2^{++})^B|0\rangle\,,$ is given by the expression which is
similar to that of the $4D, {\cal N}=2$ case \cite{GIOS}
 \bea
G^{AB}_{\tau\, (2,2)}(z_1,u_1|z_2,u_2) = -
\big(\sB^{-1}_1\big)^{AB}(\nb^+_1)^4 \d^{14}(z_1-z_2)
\d^{(-2,2)}(u_1,u_2)\,. \label{VGreen}
 \eea

The action $S_2$ (\ref{S2}) contains terms with a mixture of quantum
superfields $v^{++}$ and $q^{+}$. For further use, we diagonalize
this quadratic form by means of the special substitution of the quantum
hypermultiplet variables \footnote{A similar substitution was used
in \cite{BP07}, \cite{Kuz07} and \cite{BM15} for computing one- and
two-loop effective actions in supersymmetric theories,  and in
\cite{OV} for non-local redefinition of fields in non-supersymmetric QED.}
in the path integral \eq{Gamma0}, such that it removes the mixed
terms,
 \bea
 \label{replac}
 q^{+}_{n}(1)= h^{+}_{n}(1) - f\int d \zeta^{(-4)}_2 du_2\, G_{(1,1)}(1|2)_n{}^p
 \,i v^{++\,C}(2)\,(T^C)_p{}^{l}\, Q^{+}_{l}(2)\,,
 \eea
with $h^{+}_{n}$ being a set of new independent quantum superfields.
It is evident that the Jacobian of the variable change
(\ref{replac}) is unity. Here
$G_{\tau(1,1)}(\zeta_1,u_1|\zeta_2,u_2)_m{}^n = i\langle0| {\rm
T}{q}^{+}_{m}(\zeta_1,u_1) \tilde{q}^{+\,n}(\zeta_2,u_2)|0\rangle$
is the superfield hypermultiplet Green function in the $\tau$-frame.
This Green function is analytic with respect to both its arguments
and it satisfies the equation
 \bea
 \label{eqG}
 (\nb_1^{++})_m{}^{p}G_{\lambda\,
 (1,1)}(1|2)_p{}^n&=&\delta_m^{n}\d_A^{(3,1)}(1|2)\,.
 \eea
In $\tau$-frame the Green function can be written in the form
\bea
 G_{\tau\,(1,1)}(1|2)_m{}^n&=&(\sB_1^{-1})_m{}^{n}
 (\nb^+_1)^4(\nb^+_2)^4\f{\d^{14}(z_1-z_2)}{(u^+_1u^+_2)^3}\,.
 \label{GREEN}
  \eea
Here $\d_A^{(3,1)}(1|2)$ is the covariantly-analytic delta-function
and $(\sB)_m{}^n$ is the covariantly-analytic d'Alembertian
\eq{smile} \cite{BP15} which acts on analytic superfields $q^{+}_{m}$,
in accordance with \eq{nbRq}, as follows
  \bea
 \sB_m{}^n &=& \f12 (D^+)^4\Big\{(D^{--})^2 \delta_m^{n} + 2i(V^{--})^C (T^C)_m{}^{n} D^{--}
 +i(D^{--} V^{--})^C(T^C)_m{}^{n} \nn \\
 &&\qquad\qquad - (V^{--})^C(V^{--})^D (T^C T^D)_m{}^{n}\Big\}\,.
 \label{box3}
 \eea
Note that the covariant d'Alembertian transforms the analytic
superfields into analytic superfields.

After performing the shift \eq{replac}, the
quadratic part of the action $S_2$ \eq{S2} splits into few terms, each being bilinear in quantum superfields:
 \bea
 S_2 &=& \frac{1}{2}\int d\zeta_1^{(-4)}\,d\zeta_2^{(-4)}\,
 v_1^{++\,A}\Big\{ \sB^{AB} \d^{(3,1)}_A(1|2)
 - 2 f^2 \widetilde{Q}^{+\,m}_1 \big( T^A G_{(1,1)}T^B\big)_m{}^{n} Q^{+}_{n2}\Big\}v_2^{++\,B} \nonumber
 \\&&+ \int d\zeta^{(-4)}du\,{\bf b}^A(\nb^{++})^{2\,AB}{\bf c}^B
  +\frac{1}{2}\int d\zeta^{(-4)}du\,\varphi^A(\nb^{++})^{2\,AB}\varphi^B \nonumber \\
 &&- \int d\zeta^{(-4)}du\, \tilde{h}^{+\,m}
 (\nb^{++})_m{}^{n}h^{+}_{n}\,. \label{S22}
 \eea
Starting from the action \eq{S22} one can construct the one-loop quantum correction
$\Gamma^{(1)}[V^{++},Q^+]$ to the classical action \eq{S0}, which has
the following formal expression
 \bea
 \Gamma[V^{++},Q]&=&\frac{i}{2}\mbox{Tr}\ln\Big\{ \sB^{AB} -
 2f^2 \widetilde{Q}^{+\,m} \big( T^A G_{(1,1)}T^B\big)_m{}^{n} Q^{+}_{n}\Big\}
 -\frac{i}{2}\mbox{Tr}\ln\sB \nn \\
&& -i\mbox{Tr}\ln(\nb^{++})^2_{\rm Adj}
 +\frac{i}{2}\mbox{Tr}\ln(\nb^{++})^2_{\rm Adj}
 +i\mbox{Tr}\ln\nb^{++}_{\rm R}\,, \label{1loop}
 \eea
where subscripts ${\rm Adj}$ and ${\rm R}$ mean that the corresponding
operators are taken in the adjoint representation and that of the hypermultiplet.

The expression \eq{1loop} is the
starting point for studying the one-loop effective action in the
model \eq{S0}. In the next section we will calculate the divergent
part of \eq{1loop}. The whole dependence on the background
hypermultiplet is contained in the first term of the first line of Eq.
\eq{1loop}.

We also note that the possible structure of the one-loop
divergences in the model under consideration was discussed in
\cite{BIS} and \cite{BIMS}.

\section{Divergent part of the one-loop effective action}

The $(F^{++})^2$ part of the effective action depends only on the
background vector multiplet $V^{++}$ and is defined by the last three
terms in Eq. \eq{1loop}. More precisely,
 \bea
\Gamma_{F^2}^{(1)}[V^{++}] &=&-i\mbox{Tr}\ln(\nb^{++})^2_{\rm Adj}
 +\frac{i}{2}\mbox{Tr}\ln(\nb^{++})^2_{\rm Adj}
 +i\mbox{Tr}\ln\nb^{++}_{\rm R} \nonumber\\
&=&-i\mbox{Tr}\ln\nb^{++}_{\rm Adj}+i\mbox{Tr}\ln\nb^{++}_{\rm R}\,.
 \label{1loopV}
 \eea
Let us vary the expression \eq{1loopV} with respect to the background
gauge multiplet $(V^{++})^A\,$, keeping in mind the explicit
expressions for the covariant harmonic derivatives \eq{nbAdj} and
\eq{nbRq},
 \bea
 \d \Gamma_{F^2}^{(1)}[V^{++}] = i \Tr\, f^{ACB}\, \d
 (V^{++})^C\, G^{BA}_{(1,1)} - \Tr\,  (T^C)_m{}^{n}\, \d (V^{++})^C
 \,(G_{(1,1)})_n{}^m\,.
 \eea
Here $(G_{(1,1)})_n{}^m$ is the superfield Green function \eq{GREEN}
for operator $(\nb^{++})_n{}^{m}$ \eq{nbRq} acting on the superfields in
the representation $R$ of gauge group to which the hypermultiplet belongs. Also we denoted
$G^{BA}_{(1,1)}$  the Green function for the operator $(\nb^{++})^{BA}$
\eq{nbAdj}, which acts on superfields in adjoint representation. The
Green function $G^{BA}_{(1,1)}$ has the structure similar to \eq{GREEN},
but it is constructed in terms of the covariant
d'Alembertian \eq{box1}, \eq{Box_As_A_Sum} - \eq{Box_Rest}.

The calculation of \eq{1loopV} was discussed in details in recent
works \cite{BP15-1,BMP16,BIMS}. It is similar for abelian and non-abelian cases. Our aim is to calculate the
divergent part of the effective action \eq{1loopV}. In the
proper-time regularization scheme \cite{BP15}, \cite{BMP16}, the
divergences are associated with the pole terms of the form
$\f{1}{\varepsilon}\,$, $\varepsilon \to 0$, with $d = 6 -\varepsilon$. Taking into account the
expression for the Green functions (\ref{GREEN}), we obtain
 \bea\label{div1}
\d\Gamma^{(1)}_{F^2}[V^{++}] &=& i\int d\zeta_{1}^{(-4)} du_{1}\d
(V_1^{++})^C\Big\{f^{ACB} G^{BA}_{(1,1)}(1|2) + i(T^C)_m{}^{n}
G_{(1,1)}(1|2)_n{}^m \Big\}\Big|^{2=1}_{\rm div}.
\nonumber\\
&=& -i\int d\zeta_{1}^{(-4)} du_{1}\d (V_1^{++})^C \int_0^\infty
 d(is)(is\mu^2)^{\f\varepsilon2}  \\
&& \times\Big\{f^{ACB} (e^{is\sB_1})^{BA} + i(T^C)_m{}^{n}
(e^{is\sB_{1}})_n{}^{m}
\Big\}(\nb_1^+)^4(\nb^+_2)^4\f{\delta^{14}(z_1-z_2)}{(u^+_1u^+_2)^3}\Big|^{2=1}_{\rm
div}.\nonumber
 \eea
Here $s$ is the proper-time parameter and $\mu$ is an arbitrary
regularization parameter of mass dimension. Like in the four- and
five-dimensional cases \cite{KUZ}, one makes use of the identity
 \be\label{Delta}
(\nb^+_1)^4(\nb^+_2)^4\f{\d^{14}(z_1-z_2)}{(u^+_1u^+_2)^3}=(\nb^+_1)^4
\Big\{(u^+_1u^+_2)(\nb^-_1)^4 -(u^-_1u^+_2)\Omega_1^{--} + \sB_{1}
\f{(u^-_1u^+_2)^2}{(u^+_1u^+_2)}\Big\} \d^{14}(z_1-z_2)\,,
 \ee
where the operator $\sB$ is given by Eq. (\ref{Box_First_Part}), and
we have introduced the notation
 \bea
 \Omega^{--} = i\nb^{ab}\nb^-_a\nb^-_b - W^{-a}\nb^-_a + \frac{1}{4} (\nb^-_a
W^{- a})~. \label{O}
 \eea
To find a part of Eq. (\ref{div1}) corresponding to the first term
in Eq. (\ref{Delta}), we use the identity
\begin{equation}\label{Exponent}
e^{is\sB_1} (u_1^+ u_2^+) = e^{is\sB_1} (u_1^+ u_2^+) e^{-is\sB_1}
e^{is\sB_1}
\end{equation}
and the well-known equation
\begin{equation}
e^A B e^{-A} = B + \frac{1}{1!} [A,B] + \frac{1}{2!}[A,[A,B]] + \ldots
\end{equation}
This gives the following terms which are relevant for calculating the divergent part of the effective action:
\begin{eqnarray}
&& e^{is\sB_1} (u^+_1u^+_2) e^{-is\sB_1} \Big|^{2=1}_{\rm div} = -\f{(is)^2}{2} \Big(\nabla^M \nabla_M F^{++}
+ F^{++} (\nabla^{--} F^{++}) - \frac{1}{2}[\nabla^{--} F^{++}, F^{++}] \nonumber\\
&& + W^{+a} (\nabla^-_a F^{++})\Big) - \f{2(is)^3}{3} \nabla^M\nabla^N F^{++} \partial_M\partial_N +\ldots,\qquad
\end{eqnarray}
where dots denote terms which do not contribute to the one-loop divergences.
Adding the relevant terms coming from the expansion of the last factor in Eq. (\ref{Exponent}) we obtain
\begin{eqnarray}
&& e^{is\sB_1} (u^+_1u^+_2) \Big|^{2=1}_{\rm div} =
- \f{(is)^2}{2} \Big(\nabla^M \nabla_M F^{++} - \frac{1}{2}[\nabla^{--} F^{++}, F^{++}] + W^{+a} (\nabla^-_a F^{++})\Big)\nonumber\\
&& - \f{2(is)^3}{3} \nabla^M\nabla^N F^{++} \partial_M\partial_N +\ldots\qquad
\end{eqnarray}
In calculating a divergent part of Eq. (\ref{div1}) corresponding to
the second term of Eq. (\ref{Delta}) we can commute the exponent
with $(u_1^- u_2^+)$. After this, it is necessary to expand
$\exp(is\sB)$ in a series and keep only terms containing $(D^+)^4
(D^-)^4$. Then calculating the divergent part of the effective
action according to the standard technique, after some (rather
non-trivial) transformations we obtain the result proportional to
\begin{equation}
\nabla^M \nabla_M F^{++} + \{W^{+a}, \nabla^-_a F^{++}\} -
\frac{3}{2} [\nabla^{--} F^{++}, F^{++}] = \sB F^{++}.
\end{equation}
The fact that the operator $\sB$ appears in the final expression is
a non-trivial test of the calculation. Actually, the final
expression has the form
 \bea
 \d \Gamma^{(1)}_{F^2}[V^{++}] = \f{(C_2 - T(R))}{3 (4\pi)^3 \varepsilon}\,
\int d\zeta^{(-4)} du
\,\d V^{++ A}\, \sB F^{++ A}\,. \label{div5}
 \eea
This implies that following the same procedure as in our previous work \cite{BIMS}, it is possible to
find the action the variation of which coincides with \eq{div5}.
Up to an unessential additive constant,
 \be
 \Gamma^{(1)}_{F^2} =  \f{C_2 - T(R)}{6(4\pi)^3 \varepsilon}\,
 \int d\zeta^{(-4)} du\, (F^{++ A})^2\, =  \f{C_2 - T(R)}{3(4\pi)^3 \varepsilon}\,
 \mbox{tr}\int d\zeta^{(-4)} du\, (F^{++})^2\,, \label{div7}
 \ee
where in the last equation $F^{++} = F^{++ A} t^A$, with $t^A$ being
the generators of the fundamental representation.


The hypermultiplet-dependent part $\widetilde{Q}^+ F^{++}Q^+$ of the
one-loop counterterm comes out from the first term in \eq{1loop}. To calculate this contribution, one expands the logarithm in
the first term \eq{1loop} up to the first order and computes the
functional trace,
 \bea
 &&\frac{i}{2}\mbox{Tr}\ln\Big\{ \sB^{AB} - 2f^2
 \widetilde{Q}^{+\,m} \big( T^A G_{(1,1)}T^B\big)_m{}^{n} Q^{+}_{n}\Big\}  = \frac{i}{2}\mbox{Tr}\ln \sB \nn \\
 &&\qquad\qquad \qquad +  \frac{i}{2}\mbox{Tr}\ln\Big\{ \d^{AB} - 2f^2
 (\sB^{-1})^{AC}\widetilde{Q}^{+\,m} \big( T^C G_{(1,1)}T^B\big)_m{}^{n} Q^{+}_{n}\Big\}\,. \label{QFQ1}
 \eea
We note that, like in $4D$, $\cN=2$ SYM theory, the term
$\f{i}2\Tr\ln \sB$ does not contribute to the divergent part
\footnote{A similar analysis can be done for the contribution $\Tr
\ln \sB$ in \eq{1loop}.}. To see this, let us expose some details of
the structure of Green function for vector multiplet \eq{VGreen}. In
the limit of coincident points we need to collect eight spinorial
derivatives on delta-function $\sim(D^+)^4(D^-)^4 \d^8(\t-\t')$ in
order to obtain a non-vanishing contribution. However, the Green
function $\eq{VGreen}$ manifestly contains only four derivatives
$(D^+)^4$, while the other four spinor derivatives could be taken
from the expansion of the inverse operator $\sB$ in \eq{VGreen} up
to the fourth order in $D^-\,$. However, from this expansion we will
simultaneously gain the fourth power of the inverse flat
d'Alembertian. Thus, we will be left with the operator
$\sim\f{(D^-)^4}{\square^4}$ which can contribute only to the finite
part of effective action and so is of no interest for our
consideration.

Now, let us consider the second term in \eq{QFQ1}. Following
\cite{BIMS}, we decompose the logarithm up to the first order and
compute the functional trace
 \bea
 \Gamma^{(1)}_{QFQ} &=&  -i f^2\int d\zeta^{(-4)} du\,
 \widetilde{Q}^{+\,m}Q^{+}_{n}
 \, (\sB^{-1})^{AB } \big( T^B G_{(1,1)}T^A\big)_m{}^{n}\Big|^{2=1}_{\rm  div} \nonumber\\
 &=& -i f^2\int d\zeta^{(-4)} du\,
 \widetilde{Q}^{+\,m}Q^{+}_{n} \\
&&\quad \times \, (\sB^{-1})^{AB} \big( T^B
\sB^{-1} T^A\big)_m{}^{n}
 (u^+_1u^+_2)\, \delta^{6}(x_1-x_2)\Big|_{2=1}\,.\nonumber
  \eea
Here we made use of the explicit expression for the Green function
$(G_{(1,1)})_m{}^n$ \eq{GREEN} and once again applied the identity
\eq{Delta} for extracting the divergent contribution to effective
action. Then we decompose the inverse covariant d'Alembertians
\eq{box1} and \eq{box3} up to the second order and obtain
  \bea \Gamma^{(1)}_{QFQ} &=& -i
f^2\int d\zeta^{(-4)} du\, \widetilde{Q}^{+\,m}Q^{+}_{n} \Bigg(
\f{\d^{AB}}{\square_1} + 2f^{ACB} (F^{++})^C
\f{D_1^{--}}{\square^2_1}\Bigg) \nonumber\\
&&  \times (T^B)_m{}^{p}\Bigg( \f{\delta_p^{l}}{\square_1}
-2i(F^{++})^C (T^C)_p{}^{l} \f{D_1^{--}}{\square^2_1}\Bigg)
(T^A)_l{}^{n}(u^+_1u^+_2)
\delta^{6}(x_1-x_2)\Big|_{2=1} \nonumber\\
&=& 2i f^2 \int d\zeta^{(-4)} du\, \widetilde{Q}^{+\,m}Q^{+}_{n}(F^{++})^C
\nonumber\\
&& \times \Big\{f^{ACB} (T^B T^A)_m{}^{n} - i (T^A T^C
T^A)_m{}^{n}\Big\} \f{1}{\square^3_1}\delta^{6}(x_1-x_2)\Big|_{2=1}.
\label{ans1}
 \eea

Let is rewrite the expression within the brackets in the last line of  Eq. \eq{ans1}, using
the commutation relation
 \bea
T^C T^A = T^A T^C + i f^{CAD} T^D\,.
 \eea
Then we obtain for this expression
 \begin{equation}
 f^{ACB} T^B T^A - i T^A T^C T^A = 2 f^{ACB} T^B T^A  - i T^A T^A T^C.
 \end{equation}
Finally, we use Eq. (\ref{Casimirs}) and the identity
 \bea
 f^{ACB} T^B T^A &=& \f{i}2 f^{ACB}f^{BAD} T^D = \f{i}{2} C_{2} T^C\,,
  \eea
as well as the momentum representation of the space-time
$\delta$-function, and calculate the momentum integral in the
$\varepsilon$-regularization scheme. This leads to
 \be\f{1}{\square^3}
 \d^6(x_1-x_2)\Big|_{2=1} = \f{i}{(4\pi)^3}\f{1}{\varepsilon}\,, \quad
 \varepsilon \to 0\,.
  \ee
The result is
 \begin{equation}
\Gamma^{(1)}_{QFQ}[V^{++}, Q^+] = -\f{ 2if^2}{(4\pi)^3 \varepsilon} \int d\zeta^{(-4)}du\,
 \widetilde{Q}^{+\,m}(C_2\delta_m^{l}-C(R)_m{}^{l})(F^{++})^A\,\, (T^A)_l{}^{n}\,Q^{+}_{n}. \label{div8}
 \end{equation}

Summing up the contributions \eq{div7} and \eq{div8}, we finally
obtain the total divergent contribution
 \begin{eqnarray}
 &&\Gamma^{(1)}_{div}[V^{++}, Q^+] = \f{C_{2} - T(R)}{ 3(4\pi)^3 \varepsilon}\, \mbox{tr}\int d\zeta^{(-4)}du\,
 (F^{++})^2\nonumber\\
 &&\qquad\qquad\qquad\quad - \frac{2if^2}{(4\pi)^3\varepsilon} \int d\zeta^{(-4)} du\, {\widetilde Q}^+ (C_2-C(R)) F^{++} Q^+.
 \label{answer}
 \end{eqnarray}
We observe that the coefficients of the $(F^{++})^2$ and $\widetilde
Q^+ F^{++} Q^+$ terms in the divergent part of one-loop effective
action are proportional to the differences between the second order
Casimir operator for the adjoint representation of gauge group and
the operators $T(R)$ and $C(R)$ for the hypermultiplet
representation $R$, respectively. Since $6D$, $\cN=(1,1)$
supersymmetric Yang-Mills theory involves only the hypermultiplet in
adjoint representation of gauge group,  \eq{answer} vanishes for
this case. Hence, the $6D$, $\cN=(1,1)$ SYM theory is one-loop
finite, and there is no need to use the equations of motion
\eq{eqmF}, \eq{eqm} to prove this property.

In general, for any other choice of the irreducible representation $R$, the expression \eq{answer} does not vanish even with taking into account
the equations \eq{eqmF}, \eq{eqm}, i.e. we meet the same situation as in the abelian case considered in \cite{BIMS},
the theory is divergent already at the one-loop level \footnote{In principle, when
the hypermultiplet is in some {\it reducible} representation of gauge group, we can pick up this representation in such a way that
the coefficients before the corresponding divergent parts vanish. Such a theory will be also off-shell finite at one loop.}.
The case of pure $6D$, $\cN=(1,0)$ SYM theory corresponds to the evident choice $T(R)=0$ and $C(R)=0$ in \eq{answer},
and the one-loop divergent part is vanishing {\it on shell}, where $F^{++}=0$, in agreement with the old result of ref. \cite{HS}.


\section{Summary and outlook}

In the present paper we explicitly calculated the divergent part of the
one-loop effective action in $6D$, $\cN=(1,0)$ SYM gauge theory
coupled to the hypermultiplet in an arbitrary representation of the
gauge group. The theory was formulated in the $6D$, $\cN=(1,0)$
harmonic superspace, which preserves the manifest $6D$, $\cN=(1,0)$
supersymmetry and provides a reliable ground for conducting the
quantum field analysis.

We developed the background field quantization of the model under
consideration. Although the  $\cN=(1,0)$ SYM theories are in general anomalous\footnote{The main object of our investigation, the $\cN=(1,1)$ SYM theory, is free from anomalies.} (see, e.g., the papers \cite{FK} and references therein), the one-loop divergences can be calculated in the manifestly gauge invariant and
$\cN=(1,0)$ supersymmetric way. Anomalies are obtained by considering finite contributions and do not affect the one-loop divergences considered in this paper.
Namely, we found one-loop
divergences of the effective action both in the gauge multiplet sector
and in the hypermultiplet sector for an arbitrary gauge group and
an arbitrary hypermultiplet representation. The structure of the divergences in the
gauge multiplet sector (with all hypermultiplet contributions being suppressed)
completely matches with the results of
the analysis in refs. \cite{HS}, \cite{HS1}. In particular,
the divergences in this sector can be eliminated by a field
redefinition. This implies that the theory is {\it on-shell} finite in
the gauge multiplet sector. However, when the hypermultiplet sector
is taken into account, the situation is drastically changed. The
divergences cannot be eliminated by a field redefinition and the
theory is divergent even {\it on-shell}.

However, there is a subclass of the general theory, which deserves a
special consideration. It is the $\cN=(1,1)$ SYM theory which
includes the interacting $\cN=(1,0)$ gauge multiplet and the
$\cN=(1,0)$ hypermultiplet, both being in the same adjoint
representation. The structure of the coefficients in various terms
of the divergent part of the one-loop effective action \eq{answer}
allows us to assert that the one-loop quantum effective action of the
$\cN=(1,1)$ SYM theory does not contain the logarithmic divergences
at all, even {\it off-shell}. Such a result is entirely unexpected.

We would like to emphasize that $6D$, $\cN=(1,1)$ SYM theory is
in many aspects analogous to $4D$, $\cN=4$ SYM theory. The $4D$,
$\cN=4$ SYM theory is formulated in $\cN=2$ harmonic superspace, the
$6D$, $\cN=(1,1)$ SYM theory is formulated in $\cN=(1,0)$ harmonic
superspace, both theories include vector multiplet and
hypermultiplet in the same adjoint representation, both theories are
described by the same set of harmonics. Both theories are off-shell
finite at one loop. But $4D$, $\cN=4$ SYM theory is a completely
finite field model. Taking into account these analogies and the
results of this paper, we are led to assume that $6D$, $\cN=(1,1)$ SYM
theory can be off-shell finite at higher loops as well \footnote{It is
possible that there are new non-renormalization theorems in
$6D$, $\cN=(1,1)$ SYM theory (see the discussion of
the non-renormalization theorem in $4D$, $\cN=2$ SYM theories in
harmonic superspace approach in \cite{BKO97}, \cite{BPS}).}. The first
crucial test for such a conjecture  would be the study of the structure of the two-loop
divergences in $6D$, $\cN=(1,1)$ SYM theory. In the forthcoming paper,  we plan to carry
out an explicit calculation of the divergent part of the effective action of this theory
in the two-loop approximation.


\section*{Acknowledgements}

ILB and EAI are grateful to Kelly Stelle for useful comments and
discussion. They also thank the organizers of the MIAPP program
"Higher spin theory and duality" (May 2-27, 2016) for the
hospitality in Munich at the early stages of this work. ILB
and EAI acknowledge support from the RFBR grant, project No
15-02-06670, and from RFBR-DFG grant, project No 16-52-12012.
BSM and KVS thank Russian Science Foundation grant, project No
16-12-10306. EAI is grateful to the directorate of the Institute
of Theoretical Physics of Leibniz University of Hannover for the
hospitality at the final stage of this work.



\begin{thebibliography}{000}
\baselineskip=14pt
\parskip=0.pt
%
\bibitem{HS}
P.S. Howe, K.S. Stelle, {\it Ultraviolet Divergences in Higher
Dimensional Supersymmetric Yang-Mills Theories}, Phys. Lett. B {\bf
137} (1984) 175-180.
%
\bibitem{HS1}
P.S. Howe, K.S. Stelle, {\it Supersymmetry counterterms revisited},
Phys. Lett. B {\bf 554} (2003) 190-196, {\tt arXiv:hep-th/0211279}.
%
\bibitem{BHS}
G. Bossard, P.S. Howe, K.S. Stelle, {\it The Ultra-violet question
in maximally supersymmetric theories}, Gen. Relat. Grav. {\bf 41}
(2009) 919, {\tt arXiv:0901.4661 [hep-th]}.
%
\bibitem{BHS1}
G. Bossard, P.S. Howe, K.S. Stelle, {\it A Note on the UV behaviour
of maximally supersymmetric Yang-Mills theories}, Phys. Lett. B {\bf
682} (2009) 137-142, {\tt arXiv:0908.3883 [hep-th]}.
%
\bibitem{FT} E.S.Fradkin, A.A.Tseytlin, {\it Quantum properties of higher dimensional and
dimensionally reduced supersymmetric theories}, Nucl.Phys. B {\bf
227} (1983) 252-290. %
%
\bibitem{GIKOS}
A. S. Galperin, E. A. Ivanov, S. Kalitzin, V. I. Ogievetsky, E. S. Sokatchev, {\it Unconstrained $N{=}2$ matter, Yang-Mills and supergravity
theories in harmonic superspace},
Class. Quant. Grav. {\bf 1} (1984) 469-498 [Erratum ibid. {\bf 2} (1985) 127].
%
\bibitem{GIOS}
A. S. Galperin, E. A. Ivanov, V. I. Ogievetsky, E. S. Sokatchev,
{\it Harmonic Superspace}, Cambridge University Press, Cambridge,
2001, 306 p.
%
\bibitem{HST}
P.S. Howe, G. Sierra, P.K. Townsend, {\it Supersymmetry in six
dimensions}, Nucl. Phys. B {\bf 221} (1983) 331-348.
%
\bibitem{HSW}
P.S. Howe, K.S. Stelle, P.C. West, {\it N = 1 d = 6 harmonic
superspace}, Class. Quant. Grav. {\bf 2} (1985) 815.
%
\bibitem{Z}
B.M. Zupnik, {\it Six-dimensional supergauge theories in the
harmonic superspace}, Sov. J. Nucl. Phys. {\bf 44} (1986) 512; Yad.
Fiz. {\bf 44} (1986) 794-802.
%
\bibitem{ISZ}
E. A. Ivanov, A. V. Smilga and B. M. Zupnik, {\it Renormalizable
supersymmetric gauge theory in six dimensions},
Nucl. Phys. B {\bf 726} (2005) 131, {\tt arXiv:hep-th/0505082};\\
E.A. Ivanov, A.V. Smilga, {\it Conformal properties of
hypermultiplet actions in six dimensions}, Phys. Lett. B {\bf 637}
(2006)374, {\tt arXiv:hep-th/0510273}.
%
\bibitem{BIMS} I.L. Buchbinder, E.A. Ivanov, B.S. Merzlikin, K.V.
Stepanyantz, {\it One-loop divergences in the $6D$, $\cN= (1,0)$
abelian gauge theory}, Phys.Lett. B {\bf 763} (2016) 375-381, {\tt arXiv:1609.00975 [hep-th]}.
%
\bibitem{BIS}
G. Bossard, E. Ivanov, A. Smilga, { \it Ultraviolet behaviour of 6D
supersymmetric Yang-Mills theories and harmonic superspace}, JHEP
{\bf 1512} (2015) 085, {\tt arXiv:1509.08027 [hep-th]}.
%
\bibitem{BUCH1}
I.L. Buchbinder, E.I. Buchbinder, S.V. Kuzenko, B.A.
Ovrut, {\it The background field method for N=2 super Yang-Mills
theories in harmonic superspace}, Phys. Lett. B {\bf 417} (1998) 61,
{\tt arXiv:hep-th/9704214}.
\bibitem{BUCH2}
 E.I. Buchbinder, B.A. Ovrut, I.L. Buchbinder,
E.A. Ivanov, S.M. Kuzenko, {\it Low-energy effective action in N=2
supersymmetric field theories}, Phys. Part. Nucl. {\bf 32} (2001)
641-674; Fiz. Elem. Chast.
Atom. Yadra {\bf 32} (2001) 1222-1264;\\
I.L. Buchbinder, E.A. Ivanov, N.G. Pletnev, {\it Superfield approach
to the construction of effective action in quantum field theory with
extended supersymmetry}, Phys. Part. Nucl. {\bf 47} (2016) 291-369;
Fiz. Elem. Chast. Atom, Yadra {\bf 47} (2016) 541-698.
%
\bibitem{Kuz07} S.~M.~Kuzenko and S.~J.~Tyler, {\it Supersymmetric Euler-Heisenberg
effective action: Two-loop results}, JHEP  {\bf 0705} (2007) 081,
{\tt arXiv:hep-th/0703269}.
%
\bibitem{BM15} I.L.Buchbinder, B.S. Merzlikin, {\it On   effective
Kahler   potential   in   $\cN = 2$,   $d = 3$ SQED}, Nucl. Phys. B
{\bf 900} (2015) 80, {\tt arXiv:1505.07679 [hep-th]}.
%
\bibitem{OV} A.A. Ostrovsky, G.A. Vilkovisky, {\it The covariant
effective action in QED. One-loop magnetic moment}, J. Math. Phys.
{\bf 29} (1988) 702.
%
\bibitem{BP07} I.L. Buchbinder, N.G. Pletnev, {\it Hypermultiplet dependence of one-loop effective
action in the $\cN = 2$ superconformal theories}, JHEP {\bf 0704}
(2007) 096, {\tt arXiv:hep-th/0611145}.
%
\bibitem{Howe}
P.S. Howe, K.S. Stelle, P.K. Townsend, {\it The relaxed hypermultiplet: an unconstrained
N = 2 superfield theory}, Nucl. Phys. {\bf B} 214 (1983) 519;\\
P.S. Howe, K.S. Stelle, P.K. Townsend, {\it Miraculous ultraviolet cancelations in supersymmetry
made manifest}, Nucl. Phys. {\bf B} 236 (1984) 125.
%
\bibitem{BP15-1}I.L. Buchbinder, N.G. Pletnev, {\it Leading low-energy
effective action in the $6D$ hypermultiplet theory on a
vector/tensor background}, Phys. Lett.  B {\bf 744} (2015) 125-130,
{\tt arXiv:1502.03257 [hep-th]}.
%
\bibitem{BP15}I.L. Buchbinder, N.G. Pletnev, {\it Construction of $6D$
supersymmetric field models in ${\cal N}=(1,0)$ harmonic
superspace}, Nucl. Phys. B {\bf 892} (2015) 21-48, {\tt
arXiv:1411.1848 [hep-th]}.
%
\bibitem{BMP16}
I.L. Buchbinder, B.S. Merzlikin, N.G. Pletnev, {\it Induced
low-energy effective action in the 6D, N=(1,0) hypermultiplet theory
on the vector multiplet background}, Phys. Lett. B {\bf 759} (2016)
621-633, {\tt arXiv:1604.06186 [hep-th]}.
%
\bibitem{KUZ}
S.M. Kuzenko, I.N. McArthur, {\it  Effective action of N=4 super
Yang-Mills: N=2 superspace approach}, Phys. Lett. B {\bf 506} (2001)
140,
{\tt arXiv:hep-th/0101127}; \\
S.M. Kuzenko, I.N. McArthur, {\it Hypermultiplet effective action:
N=2 superspace approach}, Phys. Lett. B {\bf 513} (2001) 213, {\tt arXiv:hep-th/0105121}; \\
S.M. Kuzenko, I.N. McArthur, {\it  On the background field method
beyond one loop: a manifestly covariant derivative
expansion in super Yang-Mills theories}, JHEP {\bf 0305} (2003) 015, {\tt arXiv:hep-th/0302205}; \\
S.M. Kuzenko, {\it  Exact  propagators  in  harmonic  superspace},
Phys. Lett. B {\bf 600} (2004) 163, {\tt arXiv:hep-th/0407242}; \\
S.M. Kuzenko, {\it  Five-dimensional supersymmetric Chern-Simons
action as a hypermultiplet quantum correction},  Phys. Lett. B {\bf
644} (2007) 88, {\tt arXiv:hep-th/0609078}.
%
\bibitem{FK}
P.K. Townsend, G. Sierra, {\it Chiral anomalies and constraints on
the gauge group in higher-dimensional supersymmetric Yang-Mills
theories}, Nucl. Phys. B {\bf 222} (1983) 493-501;\\
S.M. Kuzenko, N. Novak, I.B. Samsonov, {\it The anomalous current
multiplet in $6D$ minimal supersymmetry}, JHEP {\bf 1602} (2016) 132, {\tt
arXiv:1511.06582 [hep-th]}.
%
\bibitem{BKO97} I.L. Buchbinder, S.M. Kuzenko, B.A. Ovrut, {\it
On the D = 4, N = 2 Non-Renormalization Theorem}, Phys.Lett. {B}
{\bf 433} (1998) 335-345, {\tt arXiv:hep-th/9710142}.
%
\bibitem{BPS}
I.L. Buchbinder, N.G. Pletnev, K.V. Stepanyantz, {\it Manifestly
${\cal N}=2$ supersymmetric regularization for ${\cal N}=2$
supersymmetric field theories}, Phys. Lett. B {\bf 75} (2015)
434-441, {\tt arXiv:1509.08055 [hep-th]}.


\end{thebibliography}
\end{document}